\documentclass{article}
\usepackage{spconf,amsmath,graphicx}
\usepackage{url}
\usepackage{cite}
\usepackage{amsmath,amssymb,amsfonts}
\usepackage{multirow, dsfont}
\usepackage{pifont,booktabs}
\usepackage{float}
\usepackage{balance}
\usepackage{graphicx, bm, array, multirow}
\usepackage{makecell}
\usepackage{dsfont}
\usepackage[ruled,vlined]{algorithm2e}

\title{Contrastive Loss Based Frame-wise Feature disentanglement for Polyphonic Sound Event Detection}

\name{Yadong Guan, Jiqing Han, Hongwei Song, Wenjie Song, Guibin Zheng, Tieran Zheng, Yongjun He 
     \thanks{This research is supported by the National Natural Science Foundation of China under Grant No.62376071.}}
\address{School of Computer Science and Technology, Harbin Institute of Technology, Harbin, China}

\begin{document}
%

\maketitle
\begin{abstract}
  Overlapping sound events are ubiquitous in real-world environments, 
  but existing end-to-end sound event detection (SED) methods still struggle to detect them effectively. 
  A critical reason is that 
  these methods represent overlapping events using shared and entangled frame-wise features, 
  which degrades the feature discrimination. 
  To solve the problem, we propose a disentangled feature learning framework 
  to learn a category-specific representation. 
  Specifically, we employ different projectors to learn the frame-wise features for each category. 
  To ensure that these feature does not contain information of other categories, 
  we maximize the common information between frame-wise features within the same category 
  and propose a frame-wise contrastive loss. 
  In addition, considering that the labeled data used by the proposed method is limited, 
  we propose a semi-supervised frame-wise contrastive loss 
  that can leverage large amounts of unlabeled data to achieve feature disentanglement. 
  The experimental results demonstrate the effectiveness of our method. 

\end{abstract}
\section{Introduction}
\label{sec:intro}

Sound event detection (SED) aims to identify the categories of sound events in an audio recording 
and corresponding onset and offset timestamps. 
It has broad applications in diverse fields, including security \cite{pandeya2020visual, nijhawan2022gun}, 
smart homes \cite{chavdar2023scarrie, khandelwal2022your}, smart cities \cite{bonet2023analysis}, 
medical industry \cite{haaland2023sound}, biodiversity detection \cite{jiang2023automatic}, etc. 
According to the criterion of whether different sound events overlap temporally, 
SED can be categorized into monophonic SED and polyphonic SED \cite{cakir2017convolutional}.
In recent years, polyphonic SED has received extensive attention 
due to the ubiquity of overlapping sounds and 
the poor performance of existing methods. 
The main challenge of polyphonic SED 
is the potential interference between overlapping event features, 
which is not conducive to learning the discriminative feature. 

To solve the problem, 
the audio separation-based SED method was proposed, 
which introduces an audio separation model as the frontend of a SED system 
to separate overlapping events and performs SED on the separated spectrogram \cite{turpault2020improving}.
However, the method requires a well-trained audio separation model, 
which is difficult to be obtained. 
Moreover, it requires reconstructing the spectrogram details of individual events, 
which is unnecessary for SED task. 

To address the limitations of separation-based approaches, 
various end-to-end methods have been proposed to detect overlapping events directly 
\cite{cakir2017convolutional, phan2022polyphonic, taslp/ChanC21, guan2023subband, li2023ast, ronchini2022benchmark}.
For example, seve-ral studies employed sophisticated models, 
such as Frequency Dynamic Convolution \cite{NamKKP22} 
and Selective Kernel Network (SK-Net) \cite{li2019selective},
to obtain more expressive features and achieved promising performance.
However, they do not explicitly consider the distinction between overlapping and non-overlapping features.
Moreover, the pattern also changes when an event is overlapped by other different events.
For this reason, some researchers treated these diverse patterns as different categories 
during the model training \cite{phan2022polyphonic}.
Neverthe-less, as the number of event categories increases, 
various overlapping patterns emerge, 
making it challenging to handle all cases.
To learn diverse overlapping patterns, 
the hard mixup method was proposed to synthesize audios with different overlapping situations \cite{C7}. 
This method was further employed to process unlabeled data 
and combined with semi-supervised methods to learn overlapping features.
The above end-to-end methods have significantly improved the detection performance. 
However, they share a common framework of 
representing overlapping events using shared and entangled frame-wise features. 
Recent research has revealed that 
the entangled features tend to move towards the classification boundary 
in the interference of different kinds of information, 
ultimately reducing feature discrimination \cite{jia2022learning}.  
Especially, if a weak sound is overlapped by other louder sounds, 
its feature discrimination will be poor.

\begin{figure}[t]
  \centering
  \centerline{\includegraphics[width=8.8cm]{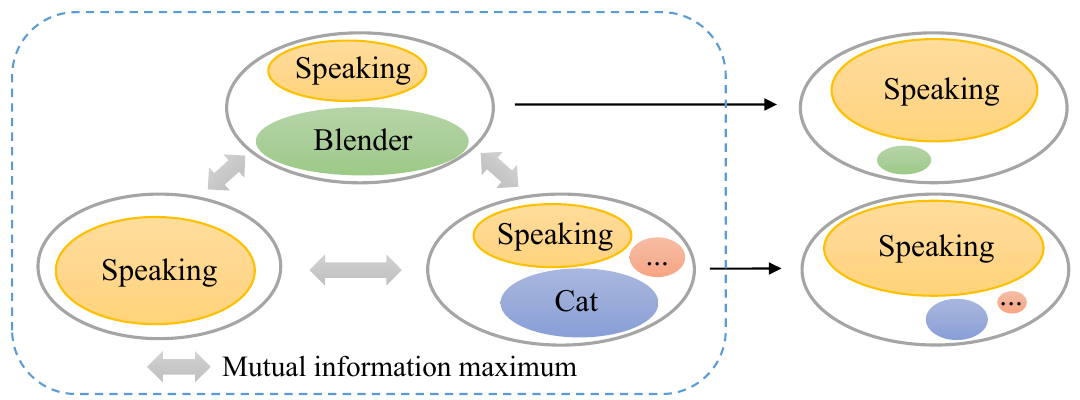}}
\caption{
Diagram of our approach. 
The circles in the dashed box are three features derived from the ``Speaking" projector. 
They all contain ``Speaking". Two of them also contain other events. 
Maximizing the mutual information of pairwise features 
will reduce irrelevant information in overlapping events, 
as shown on the right side.
}

\label{fig:res}
\end{figure}

To solve this problem, 
we introduce an idea of feature disentanglement to 
separate the information of different categories in overlapping events, 
and propose a more reasonable framework from the perspective of model design and information constraints. 
Firstly, we design multiple category-specific projectors 
to learn corresponding category information from the entangled frame-wise features. 
Then, we try to constrain the obtained features to remove the irrelevant category information. 
To achieve the goal, we consider that whether overlapped or not, 
the features of the same event always contain category-specific common information. 
Focusing on this information helps reduce other irrelevant information. 
It is the considered that the common information can be learned by mutual information maximization, 
which can be achieved by contrastive loss \cite{oord2018representation, chen2020simple}. 
Therefore, we propose a frame-wise contrastive loss that 
maximizes the MI between frame-wise features within the same category, 
as depicted in Fig. 1. 
In addition, we consider that this method requires frame-level labeled data, 
which is usually limited. 
To utilize the extra unlabeled data to further enhance the disentangling, 
we extend our method to the semi-supervised case. 
Experimental results demonstrate the effectiveness of our method. 

\section{METHODOLOGY}

\subsection{Category-specific Projector}
\label{ssec:subhead}

The overall architecture of the proposed method is shown in Fig. 2. 
We adopt the widely used Convolutional Recurrent Neural Network (CRNN) as the backbone. 
The Log-mel spectrogram $ \bm{X} \in {\mathbb{R}^{T_0 \times F}} $ is 
adopted as the input feature, where $T_0$ and $F$ represent the 
number of frames and mel-bins, respectively. 
The input feature is fed into CRNN to extract the local and temporal feature. 
Then, the obtained feature $ \bm{U} \in {\mathbb{R}^{T \times D}} $ will be mapped onto the category-specific subspaces 
using several category-specific projectors, 
\begin{equation}
  {\bm{Z}^k} = {\rm{Tanh}}\left( \bm{U} {\bm{W}^k}\right)
\end{equation}
\noindent where 
${\bm{W}^k} \in {\mathbb{R}^{D \times D/4}} $ represents a linear transformation matrix of the $k$-th class event.
$\rm{Tanh}\left( \cdot \right)$ is the activation functions. 
For ease of description, 
we refer to $ {\bm{Z}^1},...,{\bm{Z}^C} $ as the mapped features of class $ 1,...,C $. 
$C$ is the total number of event categories. 
These features are fed to their respective linear classifiers for recognition. 
Under the supervision of the ground truth, 
different transformations can learn corresponding category-specific features.

\subsection{Frame-wise Contrastive Loss}
\label{ssec:subhead}
 
After obtaining the mapped features, 
we also need to ensure that 
these features do not contain information of other irrelevant categories as much as possible.
To achieve this,  
we consider that the features of the same event always contain category-specific information, 
whether it is overlapped or not.
By maximizing the common information among them, 
the irrelevant information can be indirectly reduced. 
We use mutual information to measure common information and maximize it. 
However, mutual information is usually hard to compute for highly-dimensional features. 
Considering that in self-supervised learning methods, 
mutual information maximization can be achieved 
by minimizing the contrastive loss \cite{oord2018representation, chen2020simple}. 
Thus, we propose a frame-wise contrastive (FC) loss. 

\begin{figure}[t]
  \centering
  \centerline{\includegraphics[width=8.0cm]{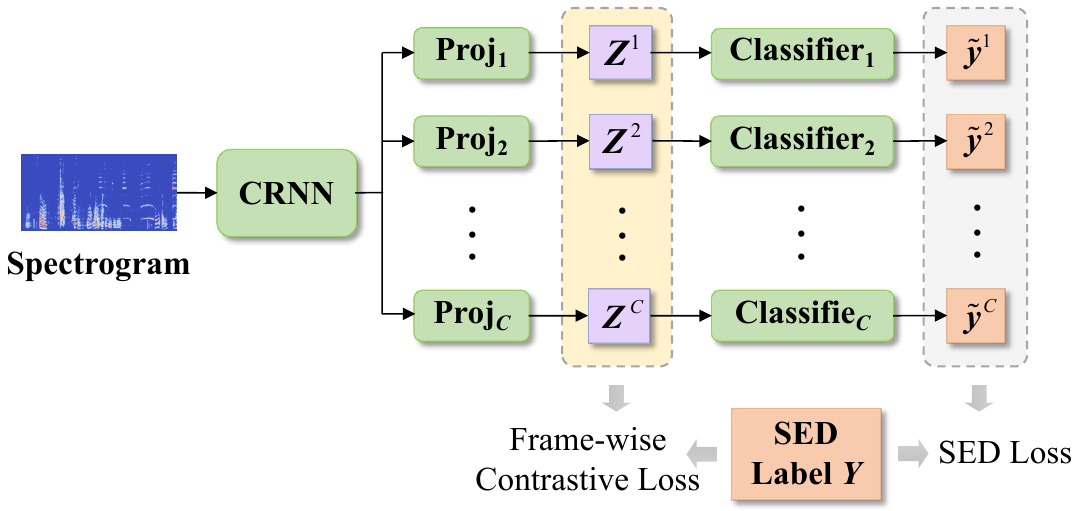}}
\caption{
   Overall architecture of the proposed method.
   The Proj denotes the category-specific projector. 
   }
\label{fig:res}
\end{figure}

We first define the anchor, the positive sample set, and the negative sample set. 
For illustration, 
taking the event $c$ as an example, 
we randomly select the feature $\bm{z}_i^c$ with a positive label of $c$-event from ${\bm{Z}^c}$, 
and set it as the anchor. 
$i$ is the time index. 
Then, for each anchor $\bm{z}_i^c$, 
we select positive samples and add them to the corresponding positive sample set $ {\mathcal{Z}_i^{c + }} $. 
The criterion of selection is that the event intersection 
between the positive sample and the anchor is only $c$. 
In other words, for the labels $y_t, y_i \in {\mathbb{R}^{C}}$ at the moment $t$ and $i$, 
if $ y_t^{\rm{T}}{{y}_i} = 1 $ and $ {{y}_t^c} = 1 $, 
we add the $ {z_t^c} $ to $ {\mathcal{Z}_i^{c + }} $. 
In addition, we add the mapped features 
whose $c$-event label is negative into the negative sample set $ {\mathcal{Z}^{c - }} $. 
In a word, for event $c$, all anchors within an audio sample share the same negative set, 
while the positive set for each anchor may be different.

For the anchor $\bm{z}_i^c$ and any element $\bm{z}_j^c$ in the $ {\mathcal{Z}_i^{c + }} $, 
we define the following contrastive loss, 
\begin{equation}
  {\cal L}\left( {\bm{z}_i^c,\bm{z}_j^c} \right){\rm{ = }} - \log \frac{{\exp \left( {\bm{z}{{_i^c}^{\rm{T}}}\bm{z}_j^c/\tau } \right)}}
  {{\sum\nolimits_{\bm{z}_k^c \in {\mathcal{Z}^{c - }}} {\exp \left( {\bm{z}{{_i^c}^{\rm{T}}}\bm{z}_k^c/\tau } \right)} }}
\end{equation}
\noindent where $\tau$ is the temperature coefficient. 
Then we compute the average loss of the anchor $\bm{z}_i^c$ and all elements in the $ {\mathcal{Z}_i^{c + }} $. 
\begin{equation}
  {\cal L}_i^c = \left\{ {\begin{array}{*{20}{l}}
  {\frac{1}{{\left| {{\mathcal{Z}_i^{c + }}} \right|}}\sum\nolimits_{\bm{z}_j^c \in {\mathcal{Z}_i^{c + }}} 
  {{\cal L}\left( {\bm{z}_i^c,\bm{z}_j^c} \right),}}&{{{\left| {{\mathcal{Z}_i^{c + }}} \right|}} \ge 1}\\
  {0,}&{\rm{else}}
  \end{array}} \right.
\end{equation}
\noindent where $ {{\left| {{\mathcal{Z}_i^{c + }}} \right|}} $ 
is the number of elements in $ {\mathcal{Z}_i^{c + }} $. 
For the $c$-th event, we apply the above loss to all anchors and average the results as follows, 
\begin{equation}
  {\cal L}^c = \left\{ {\begin{array}{*{20}{l}}
     {{\frac{1}{{{N_{y^c}}}}} \sum\limits_{i = 1}^T {\mathds{1}_{[y_i^c = 1]}}{{\cal L}_i^c},}&{N_{y^c} > 1}\\
     {0,}&{\rm{else}}
     \end{array}} \right.
\end{equation}
\noindent where ${\mathds{1}_{[\cdot]}}$ and ${N_{y^c}}$ are the indicator function 
and the number of anchors, respectively. 
$T$ is the total time dimension. 
We calculate the losses for other categories of events and get $ {\cal L}_{\rm{FC}} $, 
 \begin{equation}
  {{\cal L}_{{\rm{FC}}}} = \frac{1}{{C^+}}\sum\limits_{c = 1}^C {{\cal L}^c}
 \end{equation}
\noindent where $ {{C^+}} $ is the number of event categories occurring in the audio segment. 
$C$ is the total number of event categories.

The overall algorithm for our method is shown in Algorithm 1. 
For each category $c$, 
we first traverse the frame-level labels 
and incorporate the mapped features, whose labels are negative, into the negative sample sets. 
Subsequently, for each frame, we set mapped features with positive labels as anchors. 
Then, we select the mapped features for each anchor to generate the positive sample sets.
Finally, we compute the loss function. 
During the implementation, we employ parallelization methods to improve computational efficiency. 
\begin{algorithm}
  \caption{The calculation process of ${{\cal L}_{{\rm{FC}}}}$}\label{algorithm}
  \SetKwFunction{Union}{Union}\SetKwFunction{FindCompress}{FindCompress}
  \SetKwInOut{Input}{Input}\SetKwInOut{Output}{Output}
  \Input{
    $ {{\bm{y}}} \in {\mathbb{R}^{T \times C}} $, 
    $\bm{Z}^1,...,\bm{Z}^C \in {\mathbb{R}^{T \times D}}$, \\
  }
  \Output{$ {\cal L}_{\rm{FC}} $}

  \For {$ c = 1,2,...,C $}
  { Initialize $ \mathcal{Z}^{c-} = \varnothing $ \\
    \For {$ t = 1,2,...,T $}
      {\If {${y}_{t}^c = 0$}
        {add $ \bm{z}_{t}^c $ to $ \mathcal{Z}^{c-} $}
  		}
   \For {$ i = 1,2,...,T $}
    {\If {${y}_{i}^c = 0$}
      {break} \Else 
      {Set {$\bm{z}_{i}^c $} as anchor, initialize $ \mathcal{Z}_i^{c+} = \varnothing $ \\
      $ \bm{p} = {\bm{y}}_{i}^T {\bm{y}} $ \\
      \For {$ j = 1,2,...,T $}
        {
        \If {$p_i = 1$ and ${y}_{j}^c = 1$}
          {add $ \bm{z}_{j}^c $ to $ \mathcal{Z}_i^{c+} $}
        }
      Compute Eq.(2) and Eq.(3)
      }
      
    }
    Compute Eq.(4)
  }
  Compute Eq.(5)
  \end{algorithm}

After obtaining the ${{\cal L}_{{\rm{FC}}}}$, we add it to the original SED loss to get the final loss, 
\begin{equation}
  {\cal L} = {{\cal L}_{{\rm{SED}}}} + \lambda_1 {{\cal L}_{{\rm{FC}}}}
\end{equation}
\noindent where $ {{\cal L}_{{\rm{SED}}}} $ is the SED loss, which is the same as in \cite{shao2021rct}. 
$\lambda_1$ is the coefficient weight and we set it to 0.05.

\subsection{Semi-supervised Frame-wise Contrastive Loss}
\label{ssec:subhead}

The proposed FC loss requires frame-level labeled data, which is often limited. 
To address this problem, 
we leverage large amounts of unlabeled data 
to further improve feature learning, 
and propose a semi-supervised frame-wise contrastive (SC) loss that can utilize the unlabeled data. 
Specifically, we feed the unlabeled data into the SED model being trained and use the predictions as pseudo-labels. 
These pseudo-labels are then used to compute the loss function ${\cal L}_{{\rm{SC}}}$ 
according to the above method. 
However, the pseudo-labels may not be accurate during the initial training phase, 
thus we need to mitigate their impact. 
To this end, we weigh the ${\cal L}_{{\rm{SC}}}$ with a weighted coefficient that varies with the training epoch $t$.
The coefficient $ \lambda_2 (t) $ is defined as follows, 
\vspace{-4pt}
\begin{equation}
  \lambda_2 (t) = \left\{ {\begin{array}{*{20}{l}}
  {\lambda_1 \exp \left( { - 5{{(1 - \frac{t}{E})}^2}} \right),}&{t < E}\\
  {\lambda_1,}&{t \ge E}
  \end{array}} \right.
\end{equation}
\noindent where $E$ denotes the threshold. 
$ \lambda_2 (t) $ increases exponentially with $t$ 
and remains constant when it reaches $E$. 
We set $E$ to 100 in the experiments. 
Finally, the total loss is, 
\vspace{-4pt}
\begin{equation}
  {\cal L} = {{\cal L}_{{\rm{SED}}}} + \lambda_1 {{\cal L}_{{\rm{FC}}}} + \lambda_2 (t) {{\cal L}_{{\rm{SC}}}}
\end{equation}
\section{EXPERIMENTS}
\label{sec:typestyle}
\subsection{Experimental Setup}
\label{ssec:subhead}
We conducted the experiments using the DESED dataset in the DCASE 2021 task 4 \cite{C11}, 
which contains 10,000 frame-level labeled data, 
1,578 segment-level labeled data, and 14,412 unlabeled data. 
The duration of each audio sample is ten seconds. 
We resampled the samples at 16kHz and extracted the Log-mel spectrogram 
with a frame length and shift of 2048 and 256, respectively.  
The number of mel filters is 128. 
We employed the hard mixup, the time shifting, the pitch shifting, 
and the time masking to augment the training data \cite{shao2021rct}. 
We adopted the mainstream CRNN model in \cite{cakir2017convolutional, shao2021rct} as the baseline 
and used the Mean Teacher as the semi-supervised training strategy. 
The detailed setup can be found in \cite{shao2021rct}.
In the inference phase, 
median filtering was adopted to smooth the frame-level outputs and further prevent false positives. 
In the testing phase, we used the Validation and Public Evaluation sets, 
which we called testset 1 and testset 2. 
We adopted the PSDS1 and PSDS2 to assess the performance \cite{bilen2020framework}. 
They focus on assessing the accuracy of detecting temporal boundaries for events 
and identifying sound event categories, respectively.

\begin{figure*}[htbp]
  \centerline{\includegraphics[width=16.0cm] {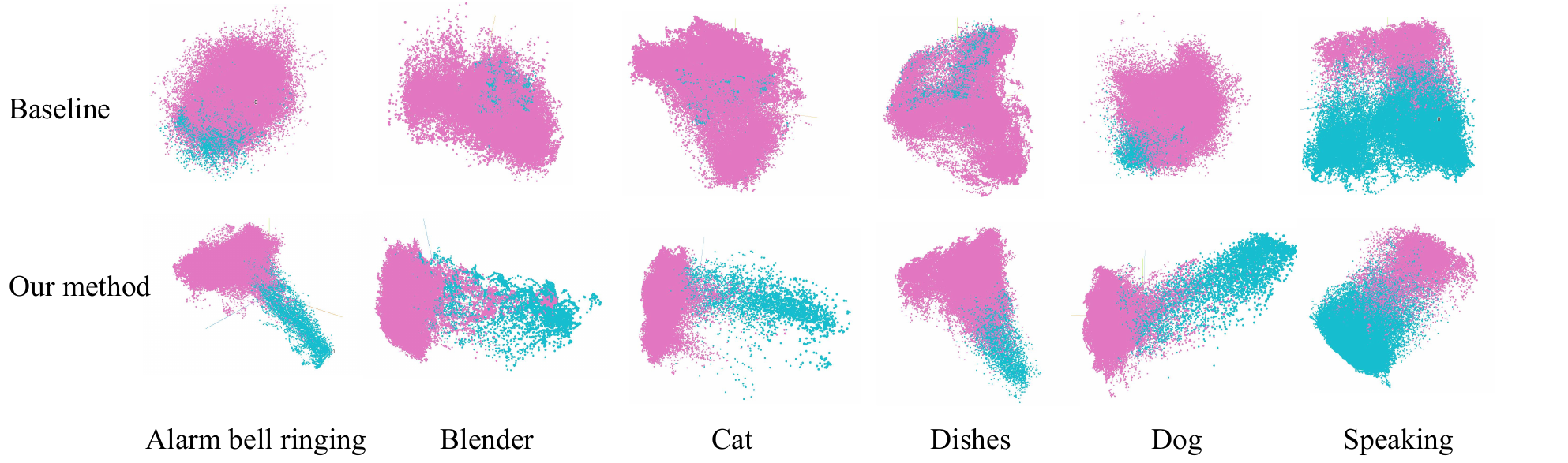}}
  \caption{
           Class-wise feature visualization of the baseline and our method. 
           Blue and pink colors indicate that the ground-truth label is ``1" and ``0", respectively.
           }
\end{figure*}

\subsection{Experimental Performance Analysis}
\label{ssec:subhead}

To demonstrate the effectiveness of our method, 
we obtain the experimental results by adding each part of our method 
and compare them with the results of the baseline, as shown in Table 1. 
Using only category-specific projectors performs worse than the baseline. 
The reason may be that after adding the projectors, 
the number of parameters increases by 36\% compared with the baseline, 
which brings the risk of over-fitting and leads to performance degradation. 
After adding the ${{\cal L}_{{\rm{FC}}}} $ on the basis of \ding{172}, 
the performance exceeds the baseline. 
Then, the performance is further improved after adding the ${{\cal L}_{{\rm{SC}}}} $. 
Compared with \ding{173}, the PSDS1 improvement of \ding{174} is relatively limited. 
The reason is that the time boundary of pseudo-labels is not accurate enough, 
which may affect the training process. 
In addition, we compare the proposed method with other common methods. 
It turns out that our method has better results on most of the metrics, especially on PSDS2. 
Notably, we only use the primitive CRNN, 
which has obvious advantages in parameter quantity.

\begin{table}[] \huge
  \centering
  \renewcommand\arraystretch{1.3}
  \caption{
    Performance comparison of different methods. 
    $ ^\ast $ denotes results from our implementation using the public codebase.
  }
  \label{tab:my-table}
  \resizebox{\columnwidth}{!}{%
  \begin{tabular}{l|ll|ll|l}
  \hline \Xhline{1.2pt}
                        & \multicolumn{2}{c|}{\textbf{Testset 1}}         & \multicolumn{2}{c|}{\textbf{Testset 2}}     &  {\multirow{2}{*}{\textbf{Para.}}}   \\ 
                        & \multicolumn{1}{l}{PSDS1 $\uparrow$ } & PSDS2 $\uparrow$  & \multicolumn{1}{l}{PSDS1 $\uparrow$ } & PSDS2 $\uparrow$  & \\ \hline \Xhline{0.5pt}
  baseline(CRNN) \cite{shao2021rct}   & \multicolumn{1}{l}{0.397}     & 0.614     & \multicolumn{1}{l}{0.437}     & 0.659                       &  1.1M  \\ 
  Projector (\ding{172})     & \multicolumn{1}{l}{0.377}     & 0.582     & \multicolumn{1}{l}{0.415}     & 0.624                                &  1.5M \\ 
  + ${{\cal L}_{{\rm{FC}}}} $ (\ding{173})         & \multicolumn{1}{l}{0.402}     & 0.624     & \multicolumn{1}{l}{\textbf{0.476}}     & 0.665    &  1.5M \\ 
  + ${{\cal L}_{{\rm{SC}}}} $ (\ding{174})   & \multicolumn{1}{l}{{0.405}}     & \textbf{0.651}     & \multicolumn{1}{l}{0.465}     & \textbf{0.696}    &  1.5M \\ \hline \Xhline{0.5pt}
  FilterAugment \cite{nam2022filteraugment}  & \multicolumn{1}{l}{0.413}     & 0.636     & \multicolumn{1}{l}{0.445}     & 0.667   &  1.0M  \\ 
  FDY-CRNN$ ^\ast  $    \cite{NamKKP22} & \multicolumn{1}{l}{\textbf{0.425}}     & 0.631     & \multicolumn{1}{l}{0.424}   & 0.646   &  11M  \\ 
  SK-CRNN$ ^\ast $ \cite{zheng2021improved} & \multicolumn{1}{l}{0.420}     & 0.625     & \multicolumn{1}{l}{0.450}    & 0.670   &  4.0M  \\ \hline \Xhline{1.2pt}
  \end{tabular}%
  }
\end{table}

To assess the effectiveness of our approach in handling overlapping and non-overlapping events, 
we divided each test set into two subsets based on the presence or absence of overlapping \cite{ESA}. 
Then, we compared the performance of our method with the baseline on these subsets, as shown in Table 2. 
In both datasets, 
the detection performance of all methods for overlapping events is much worse than for non-overlapping events.
Our method shows notable improvements compared to the baseline method, 
particularly in terms of the PSDS2 metric, where the improvement is quite remarkable.
These results suggest that our feature disentangling strategy 
exhibits a more evident improvement in the ability of event recognition. 
Furthermore, our approach demonstrates a greater improvement in detecting non-overlapping events, 
possibly due to the inherent complexity of overlapping events.

We also performed class-wise feature visualization using the principal components analysis (PCA) 
for both the baseline and our method, 
as shown in Fig. 3. 
The comparison between the two methods reveals that 
our approach yields a more compact feature distribution for each event category. 
Moreover, the features obtained by our method are farther away from the decision boundary, 
i.e., they are more discriminative.

\vspace{-5pt} 
\begin{table}[] \small
  \centering
  \renewcommand\arraystretch{1.15}
  \caption{
    Performance comparison for overlapping and non-overlapping event.
  }
  \label{tab:my-table}
  \setlength{\tabcolsep}{1.5mm}{
  \begin{tabular}{cc|cc|cc}
  \hline \Xhline{0.3pt}
                                               &          & \multicolumn{2}{c|}{\textbf{Non-overlapping}}   & \multicolumn{2}{c}{\textbf{Overlapping}}        \\ 
                                               &          & \multicolumn{1}{c}{PSDS1 $\uparrow$} & PSDS2 $\uparrow$ & \multicolumn{1}{c}{PSDS1 $\uparrow$} & PSDS2 $\uparrow$ \\ \hline
  \multicolumn{1}{c|}{\multirow{2}{*}{Testset 1}}  & Baseline & \multicolumn{1}{c}{0.465}     & 0.687     & \multicolumn{1}{c}{0.415}     & 0.672     \\ 
  \multicolumn{1}{c|}{}                        & Ours   & \multicolumn{1}{c}{\textbf{0.474}}     & \textbf{0.721}     & \multicolumn{1}{c}{\textbf{0.416}}     & \textbf{0.710}     \\ \hline
  \multicolumn{1}{c|}{\multirow{2}{*}{Testset 2}} & Baseline & \multicolumn{1}{c}{0.551}     & 0.739     & \multicolumn{1}{c}{0.351}     & 0.613     \\ 
  \multicolumn{1}{c|}{}                        & Ours   & \multicolumn{1}{c}{\textbf{0.561}}     & \textbf{0.785}     & \multicolumn{1}{c}{\textbf{0.367}}     & \textbf{0.647}     \\ \hline \Xhline{0.3pt}
  \end{tabular}%
  }
\end{table}

\section{Conclusion}
\label{sec:typestyle}

This paper proposed a contrastive loss-based frame-wise feature disentanglement method 
to separate the information of different categories in overlapping events. 
Our method utilized category-specific projectors to learn the features of each event category. 
We also proposed an FC loss that eliminates the information of other categories 
by maximizing the common information between features from different frames within the same category. 
In addition, we presented a SC loss 
that leverages large amounts of unlabeled data to facilitate further disentangling. 
The experimental results demonstrated that our method significantly enhances feature discrimination 
with a slight increase in model parameters.

\vfill\pagebreak

\bibliographystyle{IEEEbib}
\bibliography{strings,refs}

\end{document}